\documentclass[11pt]{article}
\usepackage{graphics,epsfig}
\usepackage[latin1]{inputenc}
\usepackage[english,activeacute]{babel}
\usepackage{graphicx}
\usepackage{amsmath, amsthm, amsfonts, amssymb}
\usepackage{amssymb}
\usepackage{verbatim}
\usepackage{tikz-cd}
\usepackage[all]{xy}

\newcommand{\nd}{\noindent}

\newcommand{\lra}{\leftrightarrow}

\begin{document}

\newtheorem{theo}{Theorem}[section]
\newtheorem{definition}[theo]{Definition}
\newtheorem{lem}[theo]{Lemma}
\newtheorem{prop}[theo]{Proposition}
\newtheorem{coro}[theo]{Corollary}
\newtheorem{exam}[theo]{Example}
\newtheorem{rema}[theo]{Remark}
\newtheorem{example}[theo]{Example}
\newtheorem{principle}[theo]{Principle}
\newcommand{\ninv}{\mathord{\sim}}
\newtheorem{axiom}[theo]{Axiom}

\title{Indistinguishability \textit{right from the start} in standard quantum mechanics}

\author{{\sc Federico Holik}$^{1,4}$ \ {\sc ,} \ {\sc Juan Pablo Jorge}$^1$ {\sc and} \ {\sc Cesar Massri}$^2$}

\maketitle

\begin{center}

\begin{small}
1- Instituto de F\'{\i}sica La Plata, CCT-CONICET, and Departamento
de F\'{\i}sica, Facultad de Ciencias Exactas, Universidad Nacional
de La Plata - $115$ y $49$, C.C.~$67$, $1900$ La Plata, Argentina \\

2- Departamento de Matem\'{a}tica - Facultad de Ciencias Exactas y
Naturales\\ Universidad de Buenos Aires - Pabell\'{o}n I, Ciudad
Universitaria

\end{small}
\end{center}

\vspace{1cm}

\begin{abstract}
\noindent

\end{abstract}
\bigskip
\noindent We discuss a reconstruction of standard quantum mechanics
assuming indistinguishability right from the start, by appealing to
quasi-set theory. After recalling the fundamental aspects of the
construction and introducing some improvements in the original
formulation, we extract some conclusions for the interpretation of
quantum theory.
\\
\bigskip
\begin{small}
\centerline{\em Key words: Quantum Indistinguishability - Quasi-set
theory}
\end{small}


\section{Introduction}\label{s:Introduction}

The study of collections of quantum systems has given place to great
debates in the philosophy of physics literature. The first
remarkable thing that occurs when dealing with compound quantum
systems, is that they can be prepared in entangled states
\cite{Schro1,Schro2,Bengtsson2006,werner89}. The second remarkable
thing -- and which is the subject of this work -- is that quantum
objects seem to be indistinguishable in a way that has no classical
analogue
\cite{Krause-Priscilla,Schrodinger1998,Schrodinger1952,French2006}.
These two aspects of quantum mechanics should not be confused:
indistinguishability has to be introduced as an independent axiom of
quantum theory, and quantum objects can be prepared in symmetryzed
states without showing any entanglement at all \cite{Plastino_2009}.

The status of identity of quantum systems has been an issue almost
since the conception of the theory. Many aspects of the quantum
formalism suggest that quantum systems somehow lack identity. In a
sense, they seem to be non-individuals. The positions in the
literature defer with regard to the degree with which quantum
systems depart from a classical notion of identity. Perhaps, the
most radical position was that of E. Schr\"{o}dinger, who claimed
that quantum systems were \textit{utterly indistinguishable}
\cite{Schrodinger1998,Schrodinger1952}. The connections of quantum
indistinguishability with the Principle of the Identity of
Indiscernibles (PII) has been largely debated too, most authors
arguing that it is somehow violated in the quantum domain (see for
example \cite{10.2307/20116615,10.1086/650211}). But other voices
appeared, claiming that elementary particles can be, at least,
weakly discerned
\cite{doi:10.1086/647486,10.1093/bjps/axn027,Caulton-2013,10.1093/bjps/axs038},
and tried to used that notion as an attempt to save the PII
\cite{RiseOfTheApes}. The status of identity in quantum systems has
been also discussed in terms of ontologies based in bundles of
properties (see for example \cite{daCosta2013}). The subject of
quantum indistinguishability as a whole gave place to intense
debates, and the literature about it is huge (see \cite{French2006};
see also
\cite{DBLP:journals/synthese/Arenhart12,DBLP:journals/synthese/Arenhart13,DBLP:journals/synthese/Arenhart17,DBLP:journals/jancl/KrauseA12}
and references therein). It is important to remark that, for those
who want to get stick to a classical ontology, Bohmian mechanics
\cite{Durr2009} offers an ontology of quantum objects for which --
at the metaphysical level --  there is no issue with identity:
Bohmian particles can be considered individuals that can be
identified by their \textit{hidden} trajectories in space-time. But,
as we explain in Section \ref{s:Discussion}, even in Bohmian
mechanics identity is hidden, and there is no empirical procedure
that allows to identify (and re-identify) quantum systems. This is a
remarkable feature of quantum phenomena, independently of any
interpretation: under certain circumstances, there is no way to
identify quantum systems of the same kind. Here, we call
\textit{empirical indistinguishability} to this peculiar feature of
quantum theory.

Moreover, the symmetrization postulate
\cite{Messiah-Greenberg,Girardeau-1965}, closely related to the
indistinguishability principle, can be used to explain remarkable
physical processes. Among them, one can find the Pauli exclusion
principle \cite{Marton_2013}, the Bose-Einstein condensation
\cite{Yukalov2011}, and all phenomena related to quantum statistics
in general. Currently, quantum indistinguishability is considered as
a resource, and exploited in quantum informational tasks (see for
example \cite{Adesso-Indistinguishability,LoFranco,Bellomo-2017}).
In this sense, under the light of these developments and
applications, the assumption of quantum indistinguishability is a
very powerful feature of quantum theory, that has no classical
analogue, and seems to be the right conceptual framework for the
working physicist.

Independently of the interpretation chosen or the metaphysical
commitments assumed, the symmetrization postulate and the
incapability of distinguishing quantum systems must be reflected in
the effective part of the theory (i.e., that part of the formalism
that connects with experience). At this level, the symmetrization
postulate plays a key role, in the sense that it is the mathematical
procedure that physicists found in order to give place to
predictions that describe quantum statistics correctly and, at the
same time, reflect the fact that quantum systems cannot be
discerned.

But the symmetrization postulate is implemented by appealing to a
trick. First, quantum systems are labeled in order to create a
tensor product Hilbert space, as if they were distinguishable. After
that initial labeling, quantum states are symmetrized (or
anti-symmetrized) in order to obtain the correct sates. No trace of
the initial labeling can be found in the final version of the
formalism. While mathematically correct, this procedure seems to be
rather artificial, because the initial labeling plays no real role
in the empirical predictions of the theory. Thus, many authors posed
the problem of finding a formulation of quantum theory in which
indistinguishability is taken right from the start, eliminating the
so called surplus mathematical structure of particle labeling
\cite{Post1963,Krause2003,French2006}. One candidate for solving the
``surplus structure" problem, is that of formulating quantum
mechanics using the Fock-space formalism (FSF)
\cite{Redhead1992,Redhead1991}. But, as observed in
\cite{French2006}, the FSF also makes use of particle labeling and
symmetrization in order to obtain the correct states. Among the
attempts to solve this problem, one can find \cite{Holik2008} and
\cite{Holik2010} (see also \cite{LoFranco-2018} ).

In this work, we elaborate on the proposal presented in
\cite{Holik2008} and \cite{Holik2010} (see also
\cite{Holik-PhD,Holik-Master,Holik-CompoundQuantum,Holik-EditorialAcademicaEspanola}).
We first review the fundamental features of the Fock-space
formulation of quantum mechanics and the rudiments of quasi-set
theory in Sections \ref{s:review} and \ref{s:Quasi-sets}. Next, in
Section \ref{s:Q-space}, we present the quasi-sets reformulation of
standard quantum mechanics in a new math fashion. In Section
\ref{s:Discussion}, after introducing the notion of empirical
indistinguishability, we show in which sense the assumption of
particle labelings and identities has a similar role to that of
hidden variables, playing no real role in the empirical predictions
of the theory. We argue that quantum indistinguishability is a
positive feature of quantum systems that can be treated rigorously
by using the quasi-sets formalism. This, combined with the fact that
quantum mechanics can be reformulated by assuming
indistinguishability right from the start, favours an eliminativist
approach with regard to to the notion of classical identity in the
interpretation of quantum theory. We draw our conclusions in Section
\ref{s:Conclusions}.

\section{Fock-space formalism}\label{s:review}

As is well known, the $FSF$ can be used as an alternative
mathematical framework for non-relativistic quantum mechanics
\cite{Robertson}. In order to introduce the basic ideas, in this
section we will expose the formalism in the way that is found in
most physics textbooks (see for example
\cite{Ballentine,Robertson,Greiner}). A more technical introduction
can be found in \cite{Brattelli-II,DeLaHArpe-Jones}. In standard
quantum mechanics, the Hamiltonian of $n$ identical quantum systems
with pairwise interaction, can be written as:

\begin{equation}\label{e:nhamiltonian}
H_{n}=\sum_{i=1}^{n}[(-\frac{\hbar^{2}\nabla_{i}^{
2}}{2m})+V_{1}(\mathbf{x}_{i} )+\sum_{i>j=1}^{n}
V_{2}(\mathbf{x}_{i},\mathbf{x}_{j})]
\end{equation}

\nd where we have assumed that each quantum system is subject to an
external potential $V_{1}(\mathbf{x}_{i})$ and the pairwise
interaction is represented by
$V_{2}(\mathbf{x}_{i},\mathbf{x}_{j})$. The wave function

\begin{equation}
\Psi_{n}(\mathbf{x}_{1},\ldots,\mathbf{x}_{n},t)
\end{equation}

\nd must be a solution of the Schr\"{o}dinger's equation

\begin{equation}\label{e:SchrodingerEquation}
H_{n}\Psi_{n}=i\hbar\frac{\partial}{\partial t}\Psi_{n}
\end{equation}

\nd The second quantization approach to QM has its roots in
considering equation (\ref{e:SchrodingerEquation}) as a
\emph{classical field equation}, and its solution
$\Psi_{n}(\mathbf{x}_{1},\ldots,\mathbf{x}_{n})$ as a
\emph{classical field to be quantized}. This alternative view was
originally adopted by Pascual Jordan
\cite{Schroer2011,Duncan-OnPascualJordan}, one of the foundation
fathers of quantum mechanics, and spread worldwide after the Dirac's
paper \cite{Dirac}. Furthermore, it is a standard way of describing
free fields in relativistic quantum mechanics (canonical
quantization). The space in which these quantized fields operate is
the Fock-space.

\nd The standard Fock-space is built up from the one particle
Hilbert spaces. Let $\mathcal{H}$ be a separable Hilbert space and
define:

\begin{eqnarray}
&{\mathcal{H}}^{0}&= {\mathbb{C}}
\nonumber\\
&{\mathcal{H}}^{1}&= {\mathcal{H}}
\nonumber\\
&{\mathcal{H}}^{2}&= {\mathcal{H}}\otimes{\mathcal{H}}
\nonumber\\
&\vdots&\nonumber\\
&{\mathcal{H}}^{n}&= {\mathcal{H}}\otimes \cdots \otimes
{\mathcal{H}}
\end{eqnarray}

\nd The Fock-space is thus constructed as the direct sum of $n$
particle Hilbert spaces:

\begin{equation}
{\mathcal{F}}(\mathcal{H})= \bigoplus^{\infty}_{n=0}
{\mathcal{H}}^{n}
\end{equation}

\nd When dealing with bosons or fermions, the symmetrization
postulate ($SP$) must be imposed. In order to do so, let $S_n$ be
the group of permutations of the set $\{1,2,3,\ldots,n\}$. Given any
vector $\eta=\eta_{1}\otimes\cdots\otimes \eta_{n}\in
{\mathcal{H}}^{n}$ and $P\in S_{n}$, let
$P(\eta_{1}\otimes\cdots\otimes
\eta_{n}):=\eta_{P(1)}\otimes\eta_{P(2)}\otimes\cdots\otimes
\eta_{P(n)}$. Next, define:

\begin{equation}
\sigma^{n}(\eta)=\frac{1}{n!}\sum_{P\in S_{n}}P(\eta_{1}
\otimes\cdots\otimes \eta_{n})
\end{equation}

\noindent and:

\begin{equation}
\tau^{n}(\eta)=\frac{1}{n!}\sum_{P\in S_{n}}s^{P}P(\eta_{1}
\otimes\cdots\otimes \eta_{n})
\end{equation}

\nd where:

\[ s^{P} = \left\lbrace
  \begin{array}{c l}
    1 & \text{if } P\text{ is even},\\
    -1  & \text{if } P \text{ is odd}.
  \end{array}
\right. \]

\nd Calling

\begin{equation}
{\mathcal{H}}^{n}_{\sigma}= \{\sigma^{n}(\eta)\,\,|\,\,\eta\in
{\mathcal{H}}^{n} \}
\end{equation}

\noindent and:

\begin{equation}
{\mathcal{H}}^{n}_{\tau}= \{\tau^{n}(\eta)\,\,|\,\,\eta\in
{\mathcal{H}}^{n} \}
\end{equation}

\noindent we have the Fock-space

\begin{equation}
{\mathcal{F}}^{+}(\mathcal{H})= \bigoplus _{n=0}^{\infty}
{\mathcal{H}}^{n}_{\sigma}
\end{equation}

\noindent for Bosons and

\begin{equation}
{\mathcal{F}}^{-}(\mathcal{H})= \bigoplus^{\infty}_{n=0}
{\mathcal{H}}^{n}_{\tau}
\end{equation}

\noindent for Fermions. In what follows, in order to make the
exposition clearer, we use the Dirac ket-bra notation to denote the
vectors in $\mathcal{F}(\mathcal{H})$,
$\mathcal{F}^{+}(\mathcal{H})$ and $\mathcal{F}^{-}(\mathcal{H})$.
The standard second quantization procedure considers the one
particle wave function $\psi(\mathbf{r},t)$ and its hermitian
conjugate $\psi(\mathbf{\mathbf{x},t})^{\dagger}$ as operators
acting on the Fock-space and satisfying \cite{Greiner}:

\begin{eqnarray}
&[\psi(\mathbf{x},t),\psi(\mathbf{x}',t)]_{\mp} = 0&
\nonumber\\
&[\psi(\mathbf{x},t)^{\dagger},\psi(\mathbf{x}',t)^{ \dagger}]_{\mp}
= 0&
\nonumber\\
&[\psi(\mathbf{x},t),\psi(\mathbf{x}',t)^{\dagger}]_{ \mp}=
\delta_{(\mathbf{x}-\mathbf{x}')}&
\end{eqnarray}

\noindent where $\delta({\mathbf{x}}-{\mathbf{x}'})$ is the Dirac
delta function. If $A$ and $B$ are operators, the brackets are
defined by $[A,B]_{\mp} = AB\mp BA$ (where the ``$+$" stands for
Fermions and ``$-$ for Bosons). The $n$ particle wave function
$\Psi_{n}(\mathbf{x}_{1},\ldots,\mathbf{x}_{n})$ of the standard
formulation is now written as

\begin{eqnarray}
|\psi_{n}\rangle = (n!)^{-\frac{1}{2}}\int d^{3}x_{1}\cdots\int
d^{3}x_{n}\psi(\mathbf{x}_{1})^{\dagger}
\cdots\psi(\mathbf{x}_{n})^{\dagger} |0\rangle
\Psi_{n}(\mathbf{x}_{1},\ldots,\mathbf{x}_{n})
\end{eqnarray}

\nd and it is an eigenvector (with eigenvalue $n$) of the particle
number operator:

\begin{equation}
N:=\int d^{3}x \psi(\mathbf{x})^{\dagger}\psi(\mathbf{x}).
\end{equation}

\nd We also have the relation

\begin{equation}
\Psi_{n}(\mathbf{x}_{1},\cdots,\mathbf{x}_{n}
)=(n!)^{-\frac{1}{2}}\langle
0|\psi(\mathbf{x}_{1})\cdots\psi(\mathbf{x}_{n} )|\Psi_{n}\rangle
\end{equation}

\nd An arbitrary vector of the Fock-space will be a superposition of
states with different particle number of the form

\begin{equation}\label{e:Superposition}
|\Psi\rangle=\sum_{n=0}^{\infty}\alpha_{n}|\Psi_{n}\rangle
\end{equation}

\nd \emph{and will not be in general an eigen-state of the particle
number operator}. Thus, according to the standard interpretation of
QM, its particle number will be undetermined. This is very
important, because in the presence of particle interactions, the
states may evolve into an undefined particle state like
\eqref{e:Superposition} \cite{Greiner}.

It is customary to make a Fourier decomposition of the operators
$\psi(\mathbf{x},t)$ and $\psi(\mathbf{x},t)^{\dagger}$:

\begin{subequations}
\begin{equation}
    \psi(\mathbf{x},t)=\sum_{i}u_{i}(\mathbf{x})a_{i}(t)
\end{equation}
\begin{equation}
    \psi(\mathbf{x},t)=\sum_{i}u_{i}(\mathbf{x})a_{i}^{\dagger}(t)
\end{equation}
\end{subequations}

\noindent where the complex functions $u_{i}(\mathbf{x})$ are
assumed to form a complete and orthonormal set. The operators
$a_{i}^{\dagger}(t)$ and $a_{i}^{\dagger}(t)$ aquire a simple
interpretation when the Hamiltonian is time independent. In that
case, it is useful to take the $u_{i}$'s as the eigen-functions of
the stationary Schr\"{o}dinger equation for one particle:

\begin{equation}
    \left (-\frac{\hbar^{2}}{2m}\nabla^{2}+V(\mathbf{x})\right)u_{i}(\mathbf{x})=\epsilon_{i}u_{i}(\mathbf{x})
\end{equation}

\noindent By doing so, it is possible to write
$a_{i}(t)=\exp{-\frac{i\epsilon_{i}t}{\hbar}}a_{i}$. The operators
$a_{i}^{\dagger}$ and $a_{i}$ are called creation and annihilation
operators and satisfy:

\begin{eqnarray}
&[a_{i},a_{j}]_{\mp} = \mathbf{0}&
\nonumber\\
&[a_{i}^{\dagger},a_{j}^{\dagger}]_{\mp} = \mathbf{0}&
\nonumber\\
&[a_{i},a_{j}^{\dagger}]_{ \mp}= \delta_{ij}&
\end{eqnarray}

\noindent where the ``$-$" and ``$+$" signs stand for Bosons and
Fermions, respectively.

\section{Quasi-Set Theory}\label{s:Quasi-sets}

In Zermelo-Frankel (ZF) set theory, given an object $a$, we can
always define the \emph{singleton} $A = \{a\}$. Any other object $b$
of the theory belongs to $A$, if and only if, it satisfies $b=a$.
This simple task of identification can be carried out, due to the
fact that identity is a primitive notion in the first order language
in which the theory is formulated. Quasi-set theory ($\mathfrak{Q}$
from now on) is a set-theoretical framework that can deal with
collections of truly indiscernible objects
\cite{Krause2005,Krause1992,DBLP:journals/synthese/KrauseF95,French2006,Krause2003}.
In this context, ``truly indiscernible" means that the theory is
formulated in a first order language without identity (and then,
$\mathfrak{Q}$ is an example of a \emph{non-reflexive logic}). Thus,
the most simple identification tasks that are granted in ZF, are no
longer available in $\mathfrak{Q}$. Instead of identity, an
equivalence indistinguishability relation ``$\equiv$" is postulated.
The idea behind the axioms is that quasi-sets can be used to
represent collections of quantum systems (possibly of the same kind,
and thus, indiscernible). In this framework, indiscernibility does
not implies identity: two elements of the theory might be
indiscernible (in symbols $a\equiv b$), while not being identical.
They fail to be identical in the following sense: the first order
theory of identity is not valid for all objects of the theory. In
the rest of this section, we discuss the basic formalism of
quasi-set theory.

In order to mimic elementary particles, the objects of the theory
are divided into two main groups: $m$-objects (these are
`micro-objects', intended to represent quantum systems) and
$M$-objects (`macro-objects', introduced to represent classical
objects). The $m$-objects are introduced as ur-elements in a
standard way, while quasi-sets are collections of them (or
collections of quasi-sets).

A derived notion of identity ``$=_{E}$" is introduced in the theory
($x=_{E}y$ is read '$x$ and $y$ are extensionally identical'), iff
they are both quasi-sets having the same elements (that is, $\forall
z (z \in x \lra z \in y)$), or they are both $M$-atoms and belong to
the same quasi-sets (that is, $\forall z (x \in z \lra y \in z)$).
Due to the axioms of the theory, when applied to $M$-atoms (or
collections of them), the relation of indiscernibility collapses
into that of extensional identity and has the usual properties of
the standard identity of ZFU (ZF with Urelemente). In this way, a
copy of ZFU  can then be constructed inside $\mathfrak{Q}$.

A quasi-cardinal can be assigned to quasi-sets, and it is intended
to represent how many elements they have. Using the primitive notion
of quasi-cardinal, an analogous of the axiom of weak extensionality
is postulated in $\mathfrak{Q}$. It states that those quasi-sets
that have the same quantity of elements of the same sort, are
indistinguishable. It is important to remark that collections of
$m$-atoms do no have an associated ordinal. The elements of a
quasi-set formed by $m$-objects cannot be identified by names, nor
counted, nor ordered. This is the reason why the notion of
quasi-cardinal is postulated. In \cite{Holik-2007}, the
quasi-cardinal is treated as a derived notion (see also
\cite{Holik-Master}). The problem of describing quantum systems with
undefined particle number is discussed in
\cite{Holik-2007,DBLP:journals/synthese/CostaH15,Holik-NeitherName}.

In ZF, if $w \in x$, it is easy to show that $(x - \{w \}) \cup
\{z\} = x$ if and only if $z = w$. The replacement of an element of
a set by a different element, gives place to a different set. What
happens if we try to make an analogous substitutions to a quasi-set
formed by $m$-atoms? Let $[[z]]$ denote the quasi-set with
quasi-cardinal $1$ whose only element is an indistinguishable from
$z$ (this is called the strong singleton of $z$). Suppose that $x$
is a finite quasi-set and that $z$ is an $m$-atom such that $z \in
x$. Given $w \equiv z$ and $w \notin x$, then, it is possible to
show that $(x - [[z]]) \cup [[w]] \equiv x$. This can be clearly
interpreted as follows: the permutation of indiscernibles elements
gives place to indiscernible collections.

Given $x$ and $y$, it is is possible to form $[x]$ and $[x,y]$,
which are the collections of all indiscernibles from $x$ and from
$x$ and $y$, respectively. Quasi-pairs can be built in the usual way
as: $\langle x,y\rangle:=[[x],[x,y]]$. A quasi-function $f$ is a
quasi-set formed by quasi-pairs in such a way that if $x\equiv z$
and $[[x],[x,y]]$ and $[[z],[z,w]]$ belong to $f$, we have $y\equiv
w$.

\section{The $\mathfrak{Q}$-space}\label{s:Q-space}

In this section we describe how to obtain a mathematical formalism
based in $\mathfrak{Q}$ which is equivalent to the FSF described in
Section \ref{s:review}. In this way, we provide a reformulation of
standard quantum mechanics that uses quantum indistinguishability as
a starting point, eliminating the surplus mathematical structure
discussed in \cite{Redhead1991,Redhead1992}. We follow an analogous
approach to the one given in
\cite{Holik2008,Holik2010,Holik-PhD,Holik-Master}, but introducing
technical improvements in the formulation. We use the axioms and
definitions of quasi-set theory introduced in
\cite{CriticalStudy-2005}, with minor modifications.

Using the copy of $ZFU$ in $\mathfrak{Q}$, we start by considering a
collection $\mathcal{E}= \{\epsilon_{i}\}_{i \in \mathbb{N} }$,
where $\mathbb{N}$ are the natural numbers. The $\epsilon_{i}$'s are
intended to represent outcomes of a maximal observable, and are thus
distinguishable in the classical sense. This is the reason why we
describe $\mathcal{E}$ using the classical part of $\mathfrak{Q}$.
To fix ideas, the reader can consider its elements as the possible
values of the energy of a single quantum system (assumed to have
\textit{discrete} spectrum). But it is important to keep in mind
that the $\epsilon_{i}$'s could be real numbers or just symbols used
to distinguish the different outcomes of an experiment. The only
important thing about $\mathcal{E}$ is that it is a denumerable
collection of distinguishable items (so we could have taken the
natural numbers instead of $\mathcal{E}$). The constraint in the
cardinality of $\mathcal{E}$ implies the \textit{separability}
(i.e., that it admits a denumerable basis) of the Hilbert space that
we construct, as the rigorous formulation of standard quantum
mechanics requires. The fact that $\mathcal{E}$ is denumerable and
its elements distinguishable, also implies that it can be ordered.
In the following, we choose a concrete order for its elements (given
by $\epsilon_{i}<\epsilon_{j}$, whenever $i<j$).

We want to make sense of expressions such as ``a quantum system has
energy $\epsilon_{i}$" or ``there are $n$ quanta in the energy level
$\epsilon_{i}$", ``there are $n_{i}$ quanta in the energy level
$\epsilon_{i}$ and $n_{j}$ quanta in the energy level
$\epsilon_{j}$", and so on. For this aim, we use the non-classical
part of $\mathfrak{Q}$ as follows. First, consider the quasi-set
$FIN_{\mathfrak{Q}}$ formed by all possible finite and pure
quasi-sets (with all the $m$-atoms of the same type). The existence
of $FIN_{\mathfrak{Q}}$ can be granted as follows. Assume that there
exists a quasi-set $\omega^{\lambda}$ whose quasi-cardinal is
$\aleph_{0}$ (the smallest infinite cardinal number), representing
the collection of all possible $m$-atoms of type $\lambda$. It can
be considered as an infinite and abstract reservoir a type of
quantum system collectively characterized by specifying their
charge, mass, spin, etc., represented by $\lambda$. By applying the
axiom of parts, consider $\mathcal{P}(\omega^{\lambda})$, the
quasi-set formed by its subsets. Now, apply the separation schema to
obtain
$FIN_{\mathfrak{Q}}:=\{x\in\mathcal{P}(\omega^{\lambda})\,\,|\,\,qcard(x)<\aleph_{0}\}$.
Consider next the quasi-set $\mathcal{F}$ formed by all possible
quasi-functions $f$ such that $f:\mathcal{E} \longrightarrow
FIN_{\mathfrak{Q}}$, and whenever $\langle
\epsilon_{i_{k}};x\rangle$ and $\langle \epsilon_{i_{k'}};y\rangle$
belong to $f$ and $k\neq  k'$, then $x\cap y=\emptyset$. We also
assume that the sum of the quasi-cardinals of the quasi-sets which
appear in the image of each of these quasi-functions is finite. This
means that $qcard(x)=0$ for every $x$ in the image of $f$, except
for a finite number of elements of $\mathcal{E}$. Denote by
$\mathcal{E}_{f}$ to the collection of indexes for which $f$ assigns
an $x$ such that $qcard(x)\neq 0$. There is no a piori order in
$\mathcal{E}_{f}$, but its elements can be ordered, given that they
belong to the classical part of $\mathfrak{Q}$. Each $f$ is a
quasi-set formed by ordered pairs $\langle \epsilon_{i};x\rangle$
with $\epsilon_{i}\in\mathcal{E}$ and $x\in FIN_{\mathfrak{Q}}$.
Each order pair $ \langle \epsilon_{i};x\rangle$ represents the
proposition ``the quantum number $\epsilon_{i}$ has occupation
number $qcard(x)$" or, equivalently, there are ``$qcard(x)$ quanta
in the energy level $\epsilon_{i}$". Thus, a quasi-function
$f:\mathcal{E} \longrightarrow FIN_{\mathfrak{Q}} $ can be
interpreted as ``there are $qcard(f(\epsilon_{1}))$ quanta in energy
level $\epsilon_{1}$, $qcard(f(\epsilon_{2}))$ quanta in energy
level $\epsilon_{2}$, $qcard(f(\epsilon_{3}))$ quanta in energy
level $\epsilon_{3}$ ....", in such a way that
$\sum_{i}qcard(f(\epsilon_{i}))$ is a finite number.

As discussed in Section \ref{s:Quasi-sets}, $\mathfrak{Q}$ is
constructed in such a way that the permutation of indiscernible
elements gives place to indiscernible collections. Given that these
quasi-functions described above are constructed using the
non-classical part of $\mathfrak{Q}$, the permutation of quanta has
no effect: the result of interchanging a quanta taken from a
quasi-set associated to $\epsilon_{i}$, with another one taken from
the quasi-set associated to $\epsilon_{j}$, gives place to the same
quasi-function (see also the discussion in \cite{Holik2008}). In
this way, the particle permutation operator of the standard
formalism looses its meaning here. Each $f\in\mathcal{F}$ is
characterized by the set $\mathcal{E}_{f}$ and the ``occupation
numbers" associated to each $\epsilon_{i}\in\mathcal{E}_{f}$ (given
by $qcard(f(\epsilon_{i}))$). There is no identification, nor
ordering, nor labeling of quanta in this description, because there
is no identity in the underlying logic of $\mathfrak{Q}$ for the
$m$-atoms and their collections. In this way, a quasi-function
$f\in\mathcal{F}$, faithfully represents a proposition such as
``there are $qcard(f(\epsilon_{1}))$ quanta in energy level
$\epsilon_{1}$, $qcard(f(\epsilon_{2}))$ quanta in energy level
$\epsilon_{2}$, $qcard(f(\epsilon_{3}))$ quanta in energy level
$\epsilon_{3}$ ....", without appealing to any labelling of the
quanta involved.



Now, we proceed to associate a complex vector space structure to
$\mathcal{F}$. This is a first important step if we aim to recover a
formalism equivalent to that based in Fock spaces. In order to do
that, let us recall a useful construction from algebra,
\cite{bourbaki,atiyah}. The idea of this construction is to assign
an algebraic structure (a vector space, a commutative algebra, a non
commutative algebra, etc.) to a given set, in such a way that it
\emph{generates} the desired structure. To fix ideas, let us
consider first some examples (we work the examples over the complex
numbers $\mathbb{C}$ and using standard set theory, but notice that
we can operate in a totally analogous way in $\mathfrak{Q}$).
Suppose that we have a set with one element, $S=\{x\}$, and we want
to construct the commutative algebra $A$ generated by $S$. Then, $A$
should contain expressions such as ``$1$", ``$3ix$", ``$x^2$",
``$1+ix-x^3$", and so on. If $A$ is a $\mathbb{C}$-algebra, then it
must be equal to the polynomial algebra in one variable
$\mathbb{C}[x]$. This construction is called the \textit{free
algebra generated by $S$}. The name ``\emph{free}", comes from the
fact that we are not imposing any \emph{relation} among the
generators in $S$. Another example that we can make is that of the
free associative algebra $B$ generated by $S$. If $S=\{x,y\}$, then
$B=\mathbb{C}\langle x,y\rangle$ (the notation is standard). So, for
example, $xy\neq yx$ in $B$. The last example that we want to
address is that of a complex vector space $V_{S}$ generated by
$S=\{x,y\}$. In this case, in $V$ we have expressions such as
$(3+2i)x+5iy$ or $x-y$, but the expressions $x^2$ or $xy$ are not
allowed. In fact, $x$ and $y$ become linearly independent vectors in
$V$. Thus, if we have a set $S=\{x_i\}$, then we can construct the
vector space generated by $S$ by making formal linear combinations
of its elements. By doing so, the elements of $S$ become linearly
independent. In order to construct the vector space $V$ from the set
$S$, let us denote it $V_S$, we can take the set of functions from
$S$ to $\mathbb{C}$,
\[
V_{S}=Fun(S,\mathbb{C}).
\]
It is straightforward to check that $V_{S}$ is a $\mathbb{C}$-vector
space and the indicator functions $\iota(s)$ are the canonical basis
for $V_{S}$. In fact, we can immerse $S$ inside $V_{S}$, $\iota:S\to
V_{S}$, by assigning to $s\in S$ the indicator function $\iota(s)$.
Notice that with this construction we can obtain the commutative
algebra $A$ and the non-commutative algebra $B$. For example, given
a set $S$, we can define the set of commutative monomials $S'$ and
the set of non-commutative monomials $S''$. Then, $A$ is the vector
space generated by $S'$ and $B$ by $S''$.

The above described algebraic techniques can be applied
\textit{mutatis mutandis} in $\mathfrak{Q}$, in order to generate a
complex vector space $V_{\mathcal{F}}$ using $\mathcal{F}$. Every
function $f\in\mathcal{F}$ will have a copy in $V_{\mathcal{F}}$.
This means that we will have a quasi-function $\iota:\mathcal{F}\to
V_{\mathcal{F}}$, such that $\iota(f)$ is the copy of $f$ in
$V_{\mathcal{F}}$. Thus, given $f_{1},f_{2},\cdots,f_{n}\in
\mathcal{F}$ and
$\alpha_{1},\alpha_{2},\ldots,\alpha_{n}\in\mathbb{C}$, we can form
expressions such as

\begin{equation}\label{e:LinearComb}
\alpha_{1}\iota(f_{1})+\alpha_{2}\iota(f_{2})+\cdots+\alpha_{n}\iota(f_{n})
\end{equation}

\noindent that can be interpreted as a linear combination of the
quasi-functions $f_{i}$. Thus, we can formally express superposition
states using $\mathfrak{Q}$. We will denote the elements of
$V_{\mathcal{F}}$ using Greek letters. Thus, any $\psi\in
V_{\mathcal{F}}$ is a linear combination of the form
(\ref{e:LinearComb}).

The second step that we need to give in order to recover the FSF, is
to endow the vector space $V_{\mathcal{F}}$ with a scalar product.
This can be done in different ways (see for example
\cite{Holik2008}). Here, we notice that each $f\in\mathcal{F}$ has
associated a collection of indexes $\mathcal{E}_{f}$. Given
$f,g\in\mathcal{F}$, they can only differ in the content of the sets
$\mathcal{E}_{f}$ and $\mathcal{E}_{g}$, and in the number of
quantum objects (the occupation number) associated to each value of
the observable. Thus, any physically meaningful scalar product
between $f$ and $g$, should depend only on $\mathcal{E}_{f}$,
$\mathcal{E}_{g}$, and the respective occupation numbers. Taking
into account that, for the case of indistinguishable quanta, the
order of indexes is not relevant, there is no preferred order on
$\mathcal{E}_{f}$. But it can be ordered, because its elements,
representing outcomes of experiments, are distinguishable in the
classical sense. We can exploit this to define a scalar product
without appealing to a any kind of particle labeling. With this aim,
let us first recall some useful notions of multilinear algebra. Let
$V$ be a vector space (possibly infinite-dimensional). Consider the
Tensor Algebra $T(V)$, the Symmetric Algebra $S(V)$ and the Exterior
Algebra $\bigwedge(V)$ associated to $V$. The elements of the $T(V)$
are finite sums of non-commutative monomials $v_1\otimes \ldots
\otimes v_n$ (called non-commutative polynomials), elements of
$S(V)$ are finite sums of commutative monomials $v_1\ldots v_n$
(called polynomials), and elements of $\bigwedge(V)$ are finite sums
of skew-symmetric monomials $v_1\wedge\ldots\wedge v_n$ (called
forms). For example, if $v_1$ is linearly independent from $v_2$,
then $v_1\otimes v_2-v_2\otimes v_1\neq 0$ in $T(V)$,
$v_1v_2-v_2v_1=0$ in $S(V)$ and $v_1\wedge v_2+v_2\wedge v_1=0$ in
$\bigwedge(V)$. It is important to mention that we can combine
constructions from algebra to achieve the same result. For example,
if $S$ is a set, we can take the vector space $V_{S}$ generated by
$S$, and then we can take the symmetric algebra $S(V_{S})$.
Alternatively, we can take the commutative algebra generated by $S$.
It is a standard result from algebra that both constructions agree.
Also, we can take the tensor algebra $T(V_{S})$ and the
non-commutative algebra generated by $S$. Again, both constructions
agree. For more on these constructions, see in \cite[Ch.
III]{bourbaki}.

Notice the analogy between $T(V)$, $S(V)$ and $\bigwedge(V)$ with
respect to the spaces $\mathcal{F}(\mathcal{H})$,
$\mathcal{F}_{+}(\mathcal{H})$ and $\mathcal{F}_{-}(\mathcal{H})$,
introduced in section \ref{s:review}. But the analogy should not
lead to confusion. In order to induce a scalar product in
$\mathcal{V}_{\mathcal{F}}$, we will consider formal expressions
formed by the distinguishable outcomes of an abstract observable,
without appealing to any labeling of the quantum objects involved.
Thus, consider the vector space $V_{\mathcal{E}}$ freely generated
by $\mathcal{E}$ (this means that the symbols $\epsilon_k$ and
$\epsilon_{k'}$, are now considered linearly independent vectors
whenever $k\neq k'$). First, notice that for each $f\in\mathcal{F}$,
we have en element of $V_{\mathcal{F}}$ (denote this injection by
$\iota:\mathcal{F}\longrightarrow V_{\mathcal{F}}$ ). For each pair
of these copies, we want to define a product. To do so, assign to
each quasi-function $f\in\mathcal{F}$ a non-commutative monomial in
$T(V_{\mathcal{E}})$ as follows. Specifically, given
$f\in\mathcal{F}$, consider the elements of $\mathcal{E}_{f}$
together with their respective multiplicities (given by
$qcard(f(\epsilon_{k}))$, with $\epsilon_{k}\in\mathcal{E}_{f}$).
Define the map $\Psi:\mathcal{F}\to T(V_{\mathcal{E}})$, that
assigns to $f$ the non-commutative monomial
$\epsilon_{i_{1}}\otimes\epsilon_{i_{2}}\otimes \ldots\otimes
\epsilon_{i_{m}}\in T(V_{\mathcal{E}})$, where $\epsilon_i$ appears
$k$ times if $qcard(f(\epsilon_i))=k$, and the order of the
$\epsilon_{i_{k}}$'s is chosen in such a way that $\epsilon_{i_{k}}$
appears to the left of $\epsilon_{i_{j}}$, whenever $i_{k}<i_{j}$.
For example, if for a given $f$ we have
$\mathcal{E}_{f}=\{\epsilon_{2},\epsilon_{4},\epsilon_{5}\}$, with
$qcard(f(\epsilon_{2}))=3$, $qcard(f(\epsilon_{3}))=1$ and
$qcard(f(\epsilon_{5}))=4$, and the order assigned is
$\epsilon_{2}<\epsilon_{3}<\epsilon_{5}$, the monomial assigned to
$f$ would be:
$\epsilon_{2}\otimes\epsilon_{2}\otimes\epsilon_{2}\otimes\epsilon_{3}\otimes\epsilon_{5}\otimes\epsilon_{5}\otimes\epsilon_{5}\otimes\epsilon_{5}$.

An important aspect of our map $\Psi:\mathcal{F}\to T(V)$ is that it
can be used to define a linear map $\tilde{\Psi}:V_{\mathcal{F}}\to
T(V)$, preserving the structure of addition and scalar
multiplication. Specifically, it sends the expression
$\lambda_{1}\iota(f_{1})+\lambda_{2}\iota(f_{2})+\cdots+\lambda_{n}\iota(f_{n})$
to the non-commutative polynomial
$\lambda_{1}\tilde{\Psi}(\iota(f_{1}))+\lambda_{2}\tilde{\Psi}(\iota(f_{2}))+\cdots+\lambda_{n}\tilde{\Psi}(\iota(f_{n}))\in
T(V_{\mathcal{E}})$. The above maps can be depicted as:
\[
\xymatrix{
\mathcal{F}\ar[r]^<<<<<{\Psi} \ar[d]_{\iota} & T(V_{\mathcal{E}})\\
V_{\mathcal{F}} \ar[ur]_{\tilde{\Psi}} }
\]

Using these constructions, we can induce a scalar product on
$V_{\mathcal{F}}$. Recall that if $V$ has a scalar product, then
$T(V)$, $S(V)$ and $\bigwedge(V)$ also inherit a scalar product,
\cite{bourbaki}. Hence, instead of taking $V_{\mathcal{E}}$ as the
vector space freely generated by $\mathcal{E}$, we can take
$V_{\mathcal{E}}$ as the vector space with scalar product freely
generated by $\mathcal{E}$, that is, $\epsilon_k\bot\epsilon_{k'}$
if $k\neq k'$ and $\|\epsilon_k\|=1$. Then, the map $\Psi$ assigns
to each quasi-function $f\in\mathcal{F}$ a monomial in
$T(V_{\mathcal{E}})$. Also, recall that we always have canonical
maps $\sigma:T(V_{\mathcal{E}})\to S(V_{\mathcal{E}})$ and
$\tau:T(V_{\mathcal{E}})\to\bigwedge(V_{\mathcal{E}})$ called
symmetrization and anti-symmetrization respectively. Hence, we can
assign to each $f$ a polynomial or a form, by applying the
compositions $\Psi^{+}:=\sigma\circ\Psi$ and
$\Psi^{-}:=\tau\circ\Psi$, respectively:
\begin{center}
\begin{tikzcd}
& & & S(V_{\mathcal{E}})&\\
V_{\mathcal{F}}& \mathcal{F}\arrow[l, "\iota"]\arrow[r,"\Psi"]
\arrow[urr, bend left, "\Psi^{+}"]
\arrow[drr, bend right, "\Psi^{-}"] & T(V_{\mathcal{E}})\arrow[ur, "\sigma"]\arrow[dr, "\tau"]& \\
& & & \bigwedge(V_{\mathcal{E}})
\end{tikzcd}
\end{center}

With these tools, we can define:

\begin{equation}\label{e:Pre}
    \langle \iota(f);\iota(g) \rangle_{0}:=\langle \Psi(f);\Psi(g) \rangle\,\,\forall\,\,f,g\in\mathcal{F}
\end{equation}

\begin{equation}\label{e:PreMas}
    \langle \iota(f);\iota(g) \rangle^{+}_{0}:=\langle \Psi^{+}(f);\Psi^{+}(g) \rangle\,\,\forall\,\,f,g\in\mathcal{F}
\end{equation}

\begin{equation}\label{e:PreMenos}
    \langle \iota(f),\iota(g) \rangle^{-}_{0}:= \langle \Psi^{-}(f),\Psi^{-}(g) \rangle\,\,\forall\,\,f,g\in\mathcal{F}
\end{equation}

\noindent where, in the right hand side of the above equations, we
are using the scalar products induced in $T(V_{\mathcal{E}})$,
$S(V_{\mathcal{E}})$ and $\bigwedge(V_{\mathcal{E}})$, respectively,
induced by the vector space with scalar product $V_{\mathcal{E}}$
freely generated by $\mathcal{E}$.

Clearly, \ref{e:Pre} can be extended linearly to define a complex
scalar product $\langle ...;...\rangle:V_{\mathcal{F}}\times
V_{\mathcal{F}}\to\mathbb{C}$. The completion of $V_{\mathcal{F}}$
with respect to this product gives place to a separable Hilbert
space completely equivalent to $\mathcal{F}(\mathcal{H})$. The case
of \ref{e:PreMas} is completely analogous to \ref{e:Pre}, and it can
be used to define a scalar product $\langle
...;...\rangle^{+}:V_{\mathcal{F}}\times
V_{\mathcal{F}}\to\mathbb{C}$, yielding a space completely
equivalent to $\mathcal{F}^{+}(\mathcal{H})$. Some care must be
taken with regard to \ref{e:PreMenos}: all the $f\in\mathcal{F}$
satisfying that $qcard(f(\epsilon_{i}))\geq 2$, for at least one
$i$, by construction, will have ``null norm" in the following sense:
$\langle \iota(f),\iota(f) \rangle^{-}=0$. Thus, in order to recover
a space equivalent to $\mathcal{F}^{-}(\mathcal{H})$, let us proceed
as follows. First, extend \ref{e:PreMenos} bilinearly to all
$V_{\mathcal{F}}$. Next, take the quotient space
$V_{\mathcal{F}}/\sim$, with respect to the equivalence relation
$\psi\sim \phi$, iff $\langle \psi-\phi;\psi-\phi\rangle=0$, for all
$\psi,\phi\in V_{\mathcal{F}}$ (i.e., $\psi\sim \phi$ iff their
difference has ``null norm"). With regard to the equivalence
classes, \ref{e:PreMenos} can be used to define a complex scalar
product $\langle ...;...\rangle^{-}:V_{\mathcal{F}}/\sim\times
V_{\mathcal{F}}/\sim\to\mathbb{C}$ on $V_{\mathcal{F}}/\sim$, whose
closure is formally equivalent to $\mathcal{F}^{-}(\mathcal{H})$.
For the non-symmetrized and symmetric products, we have the
following diagram:

\begin{center}
\begin{tikzcd}
               &V_{\mathcal{F}}\times V_{\mathcal{F}}\arrow[dr, "\langle ...;...\rangle"]\arrow[dl, "\langle ...;...\rangle^{+}"]\\
\mathbb{C}   & & \mathbb{C} \\
               & \iota(\mathcal{F})\times\iota(\mathcal{F}) \arrow[uu, hook]\arrow[ur, "\langle ...;...\rangle_{0}"]\arrow[ul, "\langle ...;...\rangle^{+}_{0}"]&
\end{tikzcd}
\end{center}

\noindent while for the anti-symmetric product, we have:
\begin{center}
\begin{tikzcd}
              V_{\mathcal{F}}/\sim\times V_{\mathcal{F}}/\sim\arrow[dr, "\langle ...;...\rangle^{-}"]\\
                & \mathbb{C} \\
               \iota(\mathcal{F})\times\iota(\mathcal{F}) \arrow[uu, hook]\arrow[ur, "\langle ...;...\rangle^{-}_{0}"]&
\end{tikzcd}
\end{center}

Thus, we have shown how to build, using quasi-set theory, spaces
which are equivalent to $\mathcal{F}(\mathcal{H})$,
$\mathcal{F}^{+}(\mathcal{H})$ and $\mathcal{F}^{-}(\mathcal{H})$.
This is done by defining different scalar products on
$V_\mathcal{F}$, in order to represent distinguihsable quanta
(understood as ``quanta of different kinds"), Bosons or Fermions.
The desired spaces are obtained by taking the closures with respect
to the norms induced by the scalar products of the pre-Hilbert
spaces $\langle V_\mathcal{F}, \langle ...;...\rangle \rangle$,
$\langle V_\mathcal{F}, \langle ...;...\rangle^{+} \rangle$ and
$\langle V_\mathcal{F}/\sim, \langle ...;...\rangle^{-} \rangle$,
respectively. Once this point is reached, all constructions
presented in Section \ref{s:review} can be reproduced in
$V_\mathcal{F}$ a usual, by choosing the right scalar products and
defining suitable creation and annihilation operators (see
\cite{Holik2008}). In this way, we obtain a formulation of standard
quantum mechanics by appealing to quasi-set theory, that
incorporates indistinguishability right from the start.

We now describe the action of creation and annihilation operators
for systems of Bosons and Fermions. Let us start describing Bosons.
For this case, we describe the quasi-functions constructed above by
explicitly indicating their support and occupation numbers. Thus, if
$f\in\mathcal{F}$ and
$\mathcal{E}_{f}=\{\epsilon_{i_{1}},\epsilon_{i_{3}},\epsilon_{i_{3}},\ldots,\epsilon_{i_{n}}\}$,
we denote its copy in $V_{\mathcal{F}}$ by
$\iota(f):=f_{n_{i_{1}}n_{i_{3}}n_{i_{3}}\ldots n_{i_{n}}}$, where
$n_{i_{k}}=qcard(f(\epsilon_{i_{k}}))$. It is important to remark
again that the order of the indexes has no importance in
$f_{n_{i_{1}}n_{i_{2}}n_{i_{3}}\ldots n_{i_{m}}}$: it is just a
notation that remind us the support of a given element of
$V_{\mathcal{F}}$. We could have written
$f_{n_{i_{2}}n_{i_{1}}n_{i_{3}}\ldots n_{i_{m}}}$ instead of
$f_{n_{i_{1}}n_{i_{2}}\epsilon_{i_{3}}\ldots n_{i_{m}}}$, but the
order has no physical meaning in the definition of these
quasi-functions. We also write $f_{\emptyset}\in V_{\mathcal{F}}$ to
denote the quasi-function whose $\mathcal{E}_{f}=\emptyset$ (or,
equivalently, $qcard(f_{\emptyset}(\epsilon_{i}))=0$ for all $i$),
which is intended to represent the ground state of the system under
study (i.e., a state in which all occupation numbers are zero). Let
us define the operator that creates a Boson in state $\epsilon_{k}$
by:

\begin{equation}
\mathbf{a}^{\dagger}_{\epsilon_{k}}f_{n_{i_{1}}n_{i_{2}}\ldots
n_{i_{m}}}= \sqrt{n_{k}+1}f_{n_{i_{1}}n_{i_{2}}\ldots(n_{k}+1)\ldots
n_{i_{m}}}
\end{equation}

\noindent With the above normalization, its adjoint operator
satisfies:

\begin{equation}
    \mathbf{a}_{\epsilon_{k}}f_{n_{i_1}n_{i_2}\ldots n_{i_n}}=
    \sqrt{n_{k}}f_{n_{i_1}n_{i_2}\ldots(n_{k}-1)\ldots n_{i_m}}
\end{equation}

\noindent Now, consider the action of two creation operators (with
$k< l$):

\begin{equation}\label{e:aa1}
    \mathbf{a}^{\dagger}_{\epsilon_{k}}\mathbf{a}^{\dagger}_{\epsilon_{l}}f_{n_{i_1}n_{i_2}\ldots n_{i_n}}=\sqrt{n_{k}+1}\sqrt{n_{l}+1}f_{n_{i_1}n_{i_2}\ldots (n_{k}+1)\ldots(n_{l}+1)\ldots n_{i_m}}
\end{equation}

\noindent and

\begin{equation}\label{e:aa2}
    \mathbf{a}^{\dagger}_{\epsilon_{l}}\mathbf{a}^{\dagger}_{\epsilon_{k}}f_{n_{i_1}n_{i_2}\ldots n_{i_m}}=\sqrt{n_{k}+1}\sqrt{n_{l}+1}f_{n_{i_1}n_{i_2}\ldots(n_{k}+1)\ldots (n_{l}+1)\ldots n_{i_m}}
\end{equation}

\noindent Subtracting equations \ref{e:aa1} and \ref{e:aa2}, we
obtain:

\begin{align}\label{e:aa2}
& (\mathbf{a}^{\dagger}_{\epsilon_{k}}\mathbf{a}^{\dagger}_{\epsilon_{l}}-\mathbf{a}^{\dagger}_{\epsilon_{l}}\mathbf{a}^{\dagger}_{\epsilon_{k}})f_{n_{i_1}n_{i_2}\ldots n_{i_n}}=\\
& = (\sqrt{n_{k}+1}\sqrt{n_{l}+1})\left(f_{n_{i_1}n_{i_2}\ldots
(n_{k}+1)\ldots(n_{l}+1)\ldots n_{i_m}}-f_{n_{i_1}n_{i_2}\ldots
(n_{k}+1)\ldots(n_{l}+1)\ldots n_{i_m}}\right)=\mathbf{0}
\end{align}

\noindent A similar equation holds for $k=l$. Thus, since
$f_{n_{i_1}n_{i_2}\ldots n_{i_m}}$ is arbitrary, we obtain:

\begin{equation}
    \mathbf{a}^{\dagger}_{\epsilon_{k}}\mathbf{a}^{\dagger}_{\epsilon_{l}}-\mathbf{a}^{\dagger}_{\epsilon_{l}}\mathbf{a}^{\dagger}_{\epsilon_{k}}=\mathbf{0}
\end{equation}

\noindent Similarly, it is easy to obtain:

\begin{equation}
    \mathbf{a}_{\epsilon_{k}}\mathbf{a}_{\epsilon_{l}}-\mathbf{a}_{\epsilon_{l}}\mathbf{a}_{\epsilon_{k}}=\mathbf{0}
\end{equation}

\noindent Finally, let us compute the action of
$\mathbf{a}_{\epsilon_{k}}\mathbf{a}^{\dagger}_{\epsilon_{l}}$:

\begin{equation}\label{e:aadagger1}
    \mathbf{a}_{\epsilon_{k}}\mathbf{a}^{\dagger}_{\epsilon_{l}}f_{n_{i_1}n_{i_2}\ldots n_{i_n}}=\sqrt{n_{k}-1}\sqrt{n_{l}+1}f_{n_{i_1}n_{i_2}\ldots (n_{k}-1)\ldots(n_{l}+1)\ldots n_{i_m}}
\end{equation}

\noindent For
$\mathbf{a}^{\dagger}_{\epsilon_{l}}\mathbf{a}_{\epsilon_{k}}$:

\begin{equation}\label{e:aadagger2}
    \mathbf{a}^{\dagger}_{\epsilon_{l}}\mathbf{a}_{\epsilon_{k}}f_{n_{i_1}n_{i_2}\ldots n_{i_n}}=\sqrt{n_{l}+1}\sqrt{n_{k}-1}f_{n_{i_1}n_{i_2}\ldots(n_{k}-1)\ldots(n_{l}+1)\ldots n_{i_m}}
\end{equation}

\noindent Subtracting Equations \label{e:aadagger1} and
\label{e:aadagger2}, we obtain:

\begin{equation}
    \mathbf{a}_{\epsilon_{k}}\mathbf{a}^{\dagger}_{\epsilon_{l}}-\mathbf{a}^{\dagger}_{\epsilon_{l}}\mathbf{a}_{\epsilon_{k}}=\delta_{\epsilon_{k}\epsilon_{l}}
\end{equation}

\noindent The above equations defined the desired commuting
relations. The application of the change of basis operators
\cite{Greiner} on the ground state $f_{\emptyset}$ gives

\begin{align}
& f_{\mathbf{x}\mathbf{y}}:=\frac{1}{\sqrt{2!}}\Psi(\mathbf{y})^{\dagger}\Psi(\mathbf{x})^{\dagger}f_{\emptyset}=\frac{1}{\sqrt{2}}\sum_{i}\sum_{j} u_{i}(\mathbf{x})u_{j}(\mathbf{x})\mathbf{a}^{\dagger}_{\epsilon_{i}}\mathbf{a}^{\dagger}_{\epsilon_{i}}f_{\emptyset}=\\
& =
\frac{1}{\sqrt{2}}\sum_{i}\sum_{j}u_{i}(\mathbf{x})u_{j}(\mathbf{y})f_{\epsilon_{i}\epsilon_{j}}=\frac{1}{\sqrt{2}}\sum_{i<j}
[u_{i}(\mathbf{x})u_{j}(\mathbf{y})+u_{j}(\mathbf{x})u_{i}(\mathbf{y})]f_{\epsilon_{i}\epsilon_{j}}
\end{align}

\noindent Computing the scalar product between
$f_{\epsilon_{j}\epsilon_{k}}$ and $f_{\mathbf{x}\mathbf{y}}$

\begin{equation}
 \langle f_{\epsilon_{j}\epsilon_{k}}; f_{\mathbf{x}\mathbf{y}}\rangle^{+} =   \frac{1}{\sqrt{2}}[u_{i}(\mathbf{x})u_{j}(\mathbf{y})+u_{j}(\mathbf{x})u_{i}(\mathbf{y})]
\end{equation}

\noindent we obtain the right probability amplitudes for two
particle Bosonic systems. A similar result holds for
$f_{\mathbf{x}_{1}\mathbf{x}_{2}\ldots\mathbf{x}_{N}}:=\frac{1}{\sqrt{N!}}\Psi(\mathbf{x}_{N})\ldots\Psi(\mathbf{x}_{1})^{\dagger}\Psi(\mathbf{x}_{1})^{\dagger}f_{\emptyset}$

For Fermions, let us focus on the space $\langle V_\mathcal{F}/\sim,
\langle ...;...\rangle^{-} \rangle$. A function $f\in
V_\mathcal{F}/\sim$ is characterized again by the set
$\mathcal{E}_{f}$ and its occupation numbers. But given that we are
in the quotient space, only quasi-functions whose norm is non-null
appear, and then, the occupation numbers are never greater than one.
Thus, we can use a notation in which only the occupied levels are
displayed. Thus, for $f\in V_\mathcal{F}/\sim$ such that
$\mathcal{E}_{f}=\{\epsilon_{i_1},\epsilon_{i_2},\ldots,\epsilon_{i_n}\}$,
we use the notation
$f_{\epsilon_{i_1},\epsilon_{i_2},\ldots,\epsilon_{i_n}}$. As
remarked above, while order of quantum systems has no meaning in
$f$, the set $\mathcal{E}_{f}$ can be ordered (and again, we use the
order $\epsilon_{i}<\epsilon_{j}$, iff $i<j$). Given
$\epsilon_{k}\notin \mathcal{E}_{f}$, define $s_{f,\epsilon_{k}}$ as
the minimal number of permutations needed -- starting from left to
right -- to add $\epsilon_{k}$ to the sequence
$\epsilon_{i_1}<\epsilon_{i_2}<\ldots<\epsilon_{i_n}$ in such a way
that the final result is ordered. As an example, suppose that
$\mathcal{E}_{f}=\{\epsilon_{3},\epsilon_{5},\epsilon_{7},\epsilon_{8}\}$,
and we want to add $\epsilon_{6}$ to the sequence
$\epsilon_{3}<\epsilon_{5}<\epsilon_{7}<\epsilon_{8}$, in such a way
that the final result is ordered, using the minimal number of
permutations. It is easy to check that, in this case,
$s_{f,\epsilon_{6}}=2$. If we were to add $\epsilon_{1}$, then
$n_{f,\epsilon_{1}}=0$. For $\epsilon_{9}$, we have
$n_{f,\epsilon_{9}}=4$, and so on. With these conventions in hand,
for every
$\epsilon_{k}\notin\mathcal{E}_{f_{\epsilon_{i_1}\epsilon_{i_2}\ldots\epsilon_{i_n}}}$,
define:

\begin{equation}\label{e:CreationF}
    \mathbf{c}^{\dagger}_{\epsilon_{k}}f_{\epsilon_{i_1}\epsilon_{i_2}\ldots\epsilon_{i_n}}=(-1)^{s_{f,\epsilon_{k}}}f_{\epsilon_{i_1}\epsilon_{i_2}\ldots\epsilon_{k}\ldots\epsilon_{i_n}}
\end{equation}

\noindent The interpretation of \ref{e:CreationF}, is that we have
created a quanta in level $\epsilon_{k}$. Notice also that, in
Equation \ref{e:CreationF}, $\epsilon_{k}$ is placed in between
$\epsilon_{i_{1}}$ and $\epsilon_{i_{n}}$. If, for example,
$\epsilon_{k}<\epsilon_{i_{1}}$, we should have placed
$\epsilon_{k}$ at the beginning. But this is a meaningless detail of
the notation which can also be find in the standard approach, has no
effect in the definition of the operator, and should not lead to
confusion. Naturally, whenever
$\epsilon_{k}\in\mathcal{E}_{f_{\epsilon_{i_1}\epsilon_{i_2}\ldots\epsilon_{i_n}}}$,
adding a quanta to an occupied level, should result in a null-norm
quasi-function. Since we are working in the quotient space
$V_\mathcal{F}/\sim$, we simply set (for
$\epsilon_{k}\in\mathcal{E}_{f_{\epsilon_{i_1}\epsilon_{i_2}\ldots\epsilon_{i_n}}}$):

\begin{equation}\label{e:CreationF}
    \mathbf{c}^{\dagger}_{\epsilon_{k}}f_{\epsilon_{i_1}\epsilon_{i_2}\ldots\epsilon_{i_n}}=\mathbf{0}
\end{equation}

\noindent where $\mathbf{0}$ is the neutral element for the sum in
$V_\mathcal{F}/\sim$. Now, assuming first that $k<l$, by applying
two creation operators, we obtain

\begin{equation}\label{e:CC1}
    \mathbf{c}^{\dagger}_{\epsilon_{k}}\mathbf{c}^{\dagger}_{\epsilon_{l}}f_{\epsilon_{i_1}\epsilon_{i_2}\ldots\epsilon_{i_n}}=(-1)^{s_{f,\epsilon_{k}}}(-1)^{s_{f,\epsilon_{l}}}f_{\epsilon_{i_1}\epsilon_{i_2}\ldots\epsilon_{k}\ldots\epsilon_{l}\ldots\epsilon_{i_n}}
\end{equation}

\noindent By reversing the order of application (but assuming again
$i<k$), we have

\begin{equation}\label{e:CC2}
    \mathbf{c}^{\dagger}_{\epsilon_{l}}\mathbf{c}^{\dagger}_{\epsilon_{k}}f_{\epsilon_{i_1}\epsilon_{i_2}\ldots\epsilon_{i_n}}=(-1)^{(s_{f,\epsilon_{l}}+1)}(-1)^{s_{f,\epsilon_{k}}}f_{\epsilon_{i_1}\epsilon_{i_2}\ldots\epsilon_{k}\ldots\epsilon_{l}\ldots\epsilon_{i_n}}
\end{equation}

\noindent Adding equations \ref{e:CC1} and \ref{e:CC2}, we obtain:

\begin{align}\label{e:CC2}
& (\mathbf{c}^{\dagger}_{\epsilon_{k}}\mathbf{c}^{\dagger}_{\epsilon_{l}}+\mathbf{c}^{\dagger}_{\epsilon_{l}}\mathbf{c}^{\dagger}_{\epsilon_{k}})f_{\epsilon_{i_1}\epsilon_{i_2}\ldots\epsilon_{i_n}}=\\
& =
(-1)^{n_{f,\epsilon_{k}}}(-1)^{n_{f,\epsilon_{l}}}f_{\epsilon_{i_1}\epsilon_{i_2}\ldots\epsilon_{k}\ldots\epsilon_{l}\ldots\epsilon_{i_n}}+(-1)^{(n_{f,\epsilon_{l}}+1)}(-1)^{n_{f,\epsilon_{k}}}f_{\epsilon_{i_1}\epsilon_{i_2}\ldots\epsilon_{k}\ldots\epsilon_{l}\ldots\epsilon_{i_n}}=\mathbf{0}
\end{align}

\noindent Since
$f_{\epsilon_{i_1}\epsilon_{i_2}\ldots\epsilon_{i_n}}$ is arbitrary,
we obtain:

\begin{equation}
    \mathbf{c}^{\dagger}_{\epsilon_{k}}\mathbf{c}^{\dagger}_{\epsilon_{l}}+\mathbf{c}^{\dagger}_{\epsilon_{l}}\mathbf{c}^{\dagger}_{\epsilon_{k}}=\mathbf{0}
\end{equation}

\noindent From the properties of
$\mathbf{c}^{\dagger}_{\epsilon_{k}}$, it is possible to derive
those of $\mathbf{c}_{\epsilon_{k}}$ as usual. Proceeding as usual,
we obtain:

\begin{equation}
    \mathbf{c}_{\epsilon_{k}}f_{\emptyset}=\mathbf{0}
\end{equation}

\noindent and

\begin{equation}
    \mathbf{c}_{\epsilon_{k}}\mathbf{c}_{\epsilon_{l}}+\mathbf{c}_{\epsilon_{l}}\mathbf{c}_{\epsilon_{k}}=\mathbf{0}
\end{equation}

\noindent Let us now focus on the action of
$\mathbf{c}_{\epsilon_{k}}\mathbf{c}^{\dagger}_{\epsilon_{l}}+\mathbf{c}^{\dagger}_{\epsilon_{l}}\mathbf{c}_{\epsilon_{k}}$.
Assume first that $k<l$:

\begin{equation}\label{caac1}
    \mathbf{c}_{\epsilon_{k}}\mathbf{c}^{\dagger}_{\epsilon_{l}}f_{\epsilon_{i_1}\epsilon_{i_2}\ldots\epsilon_{i_n}}= (-1)^{n_{f,\epsilon_{k}}}(-1)^{n_{f,\epsilon_{l}}}f_{\epsilon_{i_1}\epsilon_{i_2}\ldots\bar{\epsilon}_{k}\ldots\epsilon_{l}\ldots\epsilon_{i_n}}
\end{equation}

\noindent where the bar in $\bar{\epsilon}_{k}$ indicates that it
has been removed from the list (if present before).

\noindent On the other hand

\begin{equation}\label{caac2}
\mathbf{c}^{\dagger}_{\epsilon_{l}}\mathbf{c}_{\epsilon_{k}}f_{\epsilon_{i_1}\epsilon_{i_2}\ldots\epsilon_{i_n}}=
(-1)^{(n_{f,\epsilon_{l}}-1)}(-1)^{n_{f,\epsilon_{k}}}f_{\epsilon_{i_1}\epsilon_{i_2}\ldots\bar{\epsilon}_{k}\ldots\epsilon_{l}\ldots\epsilon_{i_n}}
\end{equation}

\noindent By adding Equations \ref{caac1} and \ref{caac2}, we obtain
a null result, in any case, for $k<l$. When $k=l$,
$n_{f,\epsilon_{k}}=n_{f,\epsilon_{l}}$, and we obtain an identity,
for every quasi-function. Thus, we conclude that

\begin{equation}
    \mathbf{c}_{\epsilon_{k}}\mathbf{c}^{\dagger}_{\epsilon_{l}}+\mathbf{c}^{\dagger}_{\epsilon_{l}}\mathbf{c}_{\epsilon_{k}}=\delta_{kl}
\end{equation}

\noindent The application of the change of basis operators on the
ground state $f_{\emptyset}$ gives

\begin{align}
& f_{\mathbf{x}\mathbf{y}}:=\frac{1}{\sqrt{2!}}\Psi(\mathbf{y})\Psi(\mathbf{x})f_{\emptyset}= \frac{1}{\sqrt{2}}\sum_{i}\sum_{j} u_{i}(\mathbf{x})u_{j}(\mathbf{x})\mathbf{c}^{\dagger}_{\epsilon_{i}}\mathbf{c}^{\dagger}_{\epsilon_{i}}f_{\emptyset}=\\
&
=\frac{1}{\sqrt{2}}\sum_{i}\sum_{j}u_{i}(\mathbf{x})u_{j}(\mathbf{y})f_{\epsilon_{i}\epsilon_{j}}=\frac{1}{\sqrt{2}}\sum_{i<j}
[u_{i}(\mathbf{x})u_{j}(\mathbf{y})-u_{j}(\mathbf{x})u_{i}(\mathbf{y})]f_{\epsilon_{i}\epsilon_{j}}
\end{align}

\noindent Computing the scalar product between $
f_{\epsilon_{j}\epsilon_{k}}$ and  $f_{\mathbf{x}\mathbf{y}}$, we
obtain

\begin{equation}
 \langle f_{\epsilon_{j}\epsilon_{k}}; f_{\mathbf{x}\mathbf{y}}\rangle^{-} =  \frac{1}{\sqrt{2}}[u_{i}(\mathbf{x})u_{j}(\mathbf{y})-u_{j}(\mathbf{x})u_{i}(\mathbf{y})]
\end{equation}

\noindent which yields the right probability amplitudes for two
particle Fermionic systems. Again, we obtain a similar result for
$f_{\mathbf{x}_{1}\mathbf{x}_{2}\ldots\mathbf{x}_{N}}:=\frac{1}{\sqrt{N!}}\Psi(\mathbf{x}_{N})\ldots\Psi(\mathbf{x}_{1})^{\dagger}\Psi(\mathbf{x}_{1})^{\dagger}f_{\emptyset}$.

\section{Implications of our construction}\label{s:Discussion}

So far, we have seen how to write a version of the FSF that relies
on quasi-set theory. As such, it assumes from the beginning that
quantum objects are indistinguishable. In this section, we extract
some conclusions of the above construction.

\subsection{Empirical indistinguishability and ontological concealment}\label{s:OntologicalConcealment}

There are very concrete situations in which we can isolate quantum
objects. A formidable example of this is given by electromagnetic
traps. Using such devices, it is possible to isolate atoms
\cite{AtomTraped-2010}, and even electrons \cite{ElectronTrapping}
and positrons \cite{DEHMELT1995,PositronTraped-2018}. The
researchers that trapped a positron for the first time, called it
\textit{Priscilla}. Is this labeling of a quantum object physically
meaningful? The manipulation of isolated quantum systems led some
researchers to think that they can be indeed identified. But this is
a hasty conclusion, as we explain below (see also the discussion
presented in \cite{Krause-Priscilla} and \cite{Holik-Gomez-Krause}).

The only thing that we can say for sure about a trapped atom or
electron (and quantum objects involved in similar situations), is
that there is a quantum object of a certain kind in the trap (or a
collection of them, in case there are more than one). The sentences
``there is a positron in the trap" and ``Priscilla is in the trap"
seem very similar in content. But the second one is much stronger
than the first: it says that the particular positron named Priscilla
has been trapped. It assumes that positrons can be identified and
labeled. This is not assumed in the first one: even if we cannot
identify positrons, the meaning of having just one of them in a trap
is clear. One can say that one electron leaved a track in the cloud
chamber, one photon provoked a click in the photon detector, a
positron is trapped in a ion trap, and so on. But nothing grants
that we can put physically meaningful labels to those quantum
systems. From an operational point of view, there is no reasonable
definition of particle labeling. Let us illustrate this with an
example. Suppose that two experimenters, $A$ and $B$ have two
electrons trapped in their labs. Suppose now that, by adopting the
metaphysical standpoint that electrons are full individuals, they go
on with it, and name them at time $t_{i}$. Experimenters $A$ and $B$
call their respective electrons $\alpha$ and $\beta$. The
experimenters now make their electrons interact through a scattering
process, and trap the outgoing electrons again on each lab at time
$t_{f}$. If the experiment is designed in the right way, quantum
theory \textit{predicts} that there exists no physical mechanism
-not even in principle- allowing each experimenter to know if they
have recovered their previously possessed electrons after the
scattering is performed. According to quantum theory, the identity
of each electron is gone forever, in the sense that there is no way
to tell whether $\alpha$ came back to $A$ and $\beta$ returned to
$B$, or not. The question becomes meaningless -- from an operational
standpoint -- because, according to quantum theory, it is physically
impossible to identify which is which. Of course, metaphysics can
always be put at work, and the experimenters can still assume that
after the interaction, $\alpha$ is still $\alpha$, $\beta$ is still
$\beta$, and that both electrons returned to their traps or they
where switched, even if $A$ and $B$ don't know which option has
taken place. The result of this discussion is that if we want to
assume that quantum objects can be identified, in a way strong
enough to allow naming them, this identification must be
\textit{hidden} (see also the discussion posed in
\cite{Holik-Gomez-Krause}). If assumed (at the mataphysical level),
the identity of quantum particles is as hidden as the hidden
variables of Bohmmian mechanics. An important remark is at stake
here. It doesn't matter whether the electrons \textit{actually}
undergo a scattering process as the one described above or not. The
very possibility of such an occurrence threatens their status as
individuals. The \textit{laws} governing the behavior of quantum
systems render any assumption of particle identification a purely
metaphysical claim, that cannot be subject to experimental control
in a consistent way. The only meaningful assertions about quantum
systems are of the form: ``there is one positron in that trap and an
electron outside of it", ``there are two electrons in an Helium
atom" or ``a quantum object is prepared in state $\psi$". From the
perspective of physics, any identification becomes a matter of
jargon, that can be useful in some cases (like ``Priscila is in the
trap"), but cannot be assumed to be valid in general.

\begin{figure}
\begin{center}
\includegraphics[width=11cm, height=6cm]{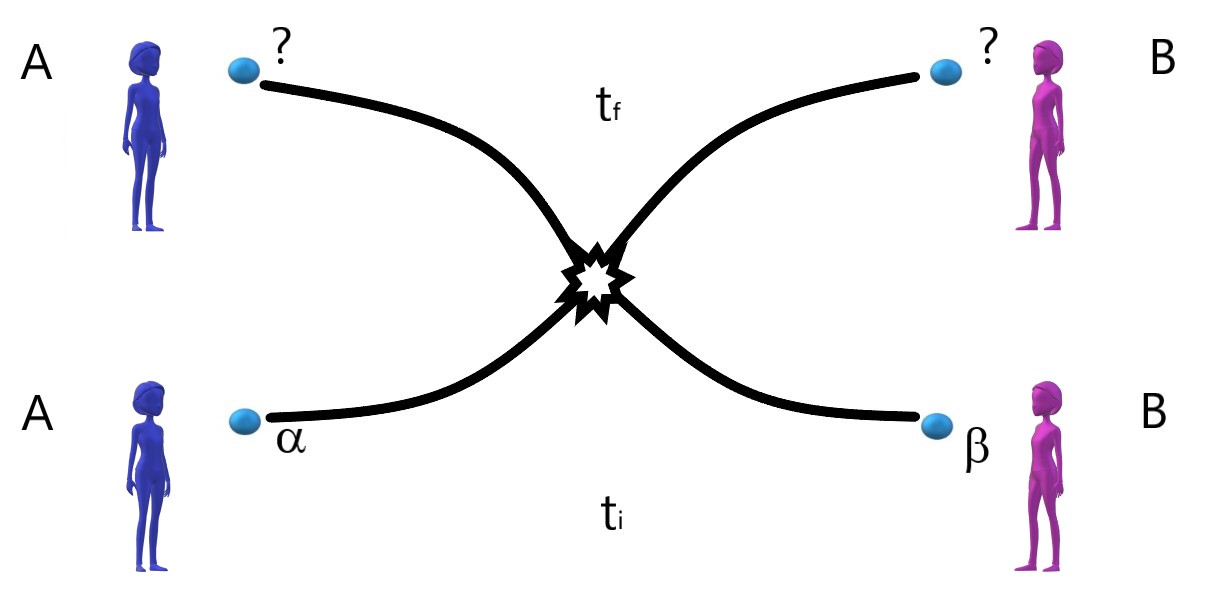}
\caption{Two experimenters, $A$ and $B$, assume that their electrons
can be named $\alpha$ and $\beta$ at time $t=t_{i}$. The electrons
undergo an interaction process. At time $t=t_{f}$, the experimenters
try to re-identify their electrons. Quantum mechanics predicts that
there exists no physical mechanism whatsoever allowing for them to
do this in a meaningful way.}
\end{center}
\end{figure}

Even if one assumes an ignorance interpretation about the identities
of the electrons, this concealment is ontological in the following
sense: if there would exist a sophisticated mechanism allowing for
the quantum objects to be identified and re-identified in an
empirically consistent way, quantum mechanics would be wrong. An
analogous situation would be to design an experiment to actually
detect simultaneously the position and momentum of a Bohmian
particle. If this were possible, we should abandon quantum mechanics
and replace it with a yet not known theory. The concealment of
variables and identities of such hidden variable theories is, in
this sense, ontological, due to the fact that it assumes that the
world is such that there exists no physical mechanism allowing us to
identify particles or detect trajectories. The world would be set as
if there would exist a conspiracy in nature, that conceals the
identities of the electrons from our experimental \textit{control}.
This situation is very different to a similar one with classical
objects. We can make an analogous experiment for classical
particles, and we may not know whether the particles were switched
or not. But we -- or some researchers from the future -- could be
able, in principle, to develop a very sophisticated mechanism
capable of determining which is which. In the classical setting,
there is no impediment in the laws of nature for knowing whether the
particles were switched or not.

The situation described above with the electrons has no classical
analogue. To see how weird this is, assume that two loving couples
$C_{1}=\{P_{1},T_{1}\}$ and $C_{2}=\{P_{2},T_{2}\}$, share two
twins, named $T_{1}$ and $T_{2}$. Suppose that $T_{1}$ and $T_{2}$
are so similar, that $C_{1}$ and $C_{2}$ do not have any means to
distinguish them physically. Now, suppose that $T_{1}$ and $T_{2}$
become involved in a strange form of interaction that hides their
identities to $P_{1}$ and $P_{2}$, the same as with electrons. When
$C_{1}$ asks to the returned twin whether he is $T_{1}$ or not, he
replies on the affirmative, and the same happens with $P_{2}$. When
asked about past memories, the twins answer that they don't remember
anything, so there is no way for $P_{1}$ and $P_{2}$ to tell whether
they were switched or not. Naturally, $P_{1}$ and $P_{2}$ want to
recover their beloved couples. $P_{1}$ wants $T_{1}$ to come back
home, and will not be willing to accept an indistinguishable from
$T_{1}$. In classical theories there always exists, no matter how
difficult it is, a mechanism allowing $P_{1}$ and $P_{2}$ to
re-identify their beloved ones, even if they can't do it in
practice. The concealment is not ontological in this case. There is
nothing in nature prohibiting the re-identification. It is just that
it is very difficult for $C_{1}$ and $C_{2}$ to do it.

The situation is much worse than not knowing. By itself, concealment
is not a problem for physics. There are many physical quantities
which are not directly observable. For example, we don't observe an
electric field. We just observe the action of it on some test
particle that we put in a certain space region. But the subtle point
here is that we can control, at lest in principle, electric fields
in such a way to make reasonable predictions in the lab. We usually
postulate entities that we cannot directly observe, but we can
\textit{control} them in such a way that their effects can be
expressed in empirically testable mathematical laws. Even if doing
this is too difficult for many complex systems, we know that,
according to the classical description of physics, there is no
intrinsic concealment in nature, and that we could do it in
principle. It may also happen that we are not able to do certain
things at a stage of development of a theory, but we can hope for
doing them in the future, without invalidating the theory. A
paradigmatic example of this is the manipulation of a single atom.
No one dreamed about that possibility in the early days of quantum
mechanics, but according to the theory, there were nothing in nature
preventing us to do so. It is perfectly possible to trap one atom or
one electron using electromagnetic devises. The situation is very
different for the case of indistinguishable particles: if there were
a possibility of identifying (and re-identifying) in a consistent
and verifiable way, quantum mechanics -- in its current form --
would be simply wrong.

The real problem with postulating hidden variables and identities in
quantum mechanics is that, even if we assume hidden identities or
trajectories, there is no physical experiment whatsoever (and it
will never will if quantum mechanics is true), that allow us to
re-identify quantum objects or to control the values of the hidden
variables. In this sense, there is a very clear empirical feature of
quantum phenomena that indicates that quantum systems cannot be
re-identified. We may calle this feature the \textit{ principle of
no re-identification}: if the experimenters adopt the metaphysical
assumption that quantum objects can be labeled at a given moment, in
the general case, there is no physical procedure allowing them to
re-identify those systems in the future. It is important to stress
that these limitations on the possibility of identifying
consistently -- from an empirical standpoint -- are a common feature
of \textit{all} interpretations which are compatible with the
predictions of quantum theory. And this is not just a metaphysical
assumption: it is a constitutive physical feature of how the world
is according to quantum theory. Remarkably enough, it is consistent
with the laboratory observations up to now.

\subsection{Lack of something?}\label{s:LackOfSomething}

The operational impossibility of re-identification described above
is traditionally interpreted as a negative feature of quantum
systems. Taking indistinguishability seriously from an ontological
standpoint is usually criticized as a weird move, in favor of a
classical ontology based in standard individuals. The ``in" in
``indistinguishability", the ``non" in ``non-individuals" and the
``weak" in ``weakly discernible objects", usually have a negative
connotation. It is as if quantum objects would lack of something,
because of the impossibility of naming or identifying them in a
consistent way. The situation is usually presented as if being an
individual in the classical sense would be ``metaphysically"
stronger than being a non-individual. But this picture is incorrect,
if we stay close to the actual development of physics.

The theoretical and experimental research of the last decades,
remarkably enhanced by the development of quantum technologies,
indicates that, rather than a weakness, non-individuality is a
positive feature of quantum systems. Put in simple words:
non-individuals can do things that classical entities obeying the
classical theory of identity, cannot do.

Indeed, quantum indistinguishability lies at the basis of many
quantum technologies. for example, the Hong-Ou-Mandel
\cite{Hong-Ou-Mandel} effect relies heavily in the fact the photons
are prepared in a state in which they cannot be individuated by any
means. The more indistinguishable the photons are prepared, more
clearly the interference pick reveals itself in the laboratory. This
device can be used to measure time differences with a very high
precision. Recently, quantum indistinguishability has been
identified as a resource for generating quantum entanglement
\cite{LoFranco,Bellomo-2017,Adesso-Indistinguishability}, one of the
main ingredients of quantum technologies. Boson sampling
\cite{BosonSampling-Introduction}, is a very simple one-way quantum
computer that is of big interest for testing fundamental features of
quantum computing. It relies solely on indistinguishable photons
(and remarkably, no entanglement seems to be involved). Recently,
the notion of indistinguishability has been studied in connection
with quantum contextuality
\cite{DBLP:journals/entropy/BarrosHK17,de_barros_indistinguishability_2019},
showing its potential as an ontological principle. The main point
that we want to make here is that the physical properties of quantum
systems with regard to identity are behind many relevant
developments in physics that have no analogue in the classical
domain. The principle of indistinguishability -- closely related to
the principle of no re-identifications describe above -- is one of
the strongest features of quantum mechanics and is used by working
physicists as the correct tool to describe nature.

Thus, if indistinguishability is so important for predicting new
physical phenomena and for the development of technological devises,
why not taking it seriously at the ontological level? The fact that
we can always postulate hidden identities, should not lead to
confusion: the impossibility of distinguishing predicted by quantum
theory reveals a positive feature of quantum entities, and can be
used as a resource in quantum information theory. To postulate the
existence of weak discernible entities or hidden identities is of no
use for the problems that the quantum physicists need to deal with.
The explanatory power of these notions is usually empty (when not
misleading) for the working physicist, and are dispensable.

Is it possible to make sense of non-individuals? Quasi-set theory
shows that this is perfectly possible: we have a formal framework
(at least one) that allows to speak about non-individuals in a
rigorous way. This formalism gives a rigorous ground to the ideas
that many physicists used in order to explain and predict new
phenomena. Furthermore, the content of Section \ref{s:Q-space},
shows that it is even possible to \textit{reformulate} quantum
mechanics using quasi-set theory. In this way, we have exposed a
formulation of quantum mechanics in which quantum objects are
considered as non-individuals \textit{right from the start}.

These developments do not imply that the ontologies based in
individual entities should be rejected in quantum mechanics. But our
work shows that the identity of quantum objects can be dispensed
with (in the sense of being an \textit{eliminable} feature). This
eliminativist move works in different levels. First, it works in the
operational level, given that no ontology based on individuals can
make predictions in which quantum objects can be identified and
re-identified in a satisfactory way. We have seen that this is not
possible for quantum systems -- provided that quantum mechanics is
assumed to be correct. On the other hand, we have shown that it is
possible -- and useful -- to consider non-individuality as a
positive feature of quantum systems (and even as a
\textit{resource}), that can be formally described in a mathematics
which relies in quasi-set theory.

\section{Conclusions}\label{s:Conclusions}

In this work we have reviewed the approach to standard quantum
mechanics based in quasi-set theory. We have shown how to recover a
Fock-space formalism based in quasi-set theory, accomplishing the
task of reformulating quantum mechanics using non-individuality
right from the start.

We have also elaborated on \textit{indistinguishability} from an
operational standpoint, relating it to the limitations that quantum
theory imposes on the attemtps to identify and re-identify quantum
objects (suggesting that a \textit{no re-identification principle}
is operating in the quantum domain). We have argued that this
peculiar feature of quantum systems, which must be necessarily
shared by all interpretations which are consistent with the
predictions of quantum theory, leads to a sort of
\textit{ontological concealment} of identity in those
interpretations which are based on individuals (such as Bohmian
mechanics). We have argued that this, together with the formulation
of quantum theory described in Section \ref{s:Q-space}, makes the
notion of identity eliminable, in the sense that it can be dispensed
with. We have argued that, far from being a negative notion (``a
lack of something"), non-individuality is a positive feature of
quantum systems that can be rigorously formalized using quasi-set
theory. It can also be used as a predictive tool and as a resource
in quantum information theory, and it opens the door to a sound
formulation of quantum mechanics based in non-individuals.

\vskip1truecm

\noindent {\bf Acknowledgements}

\noindent This work was partially supported by FONCYT, Universidad
Austral and the grant ``Per un'estensione semantica della Logica
Computazionale Quantistica- Impatto teorico e ricadute
implementative", RASSR40341.

\end{document}